# Unexpected Anisotropic Two Dimensional Electron Gas at the LaAlO$_3$/SrTiO$_3$ (110) Interface

A. Annadi[1,2], X. Wang[1,2], K. Gopinadhan[1,3], W. M. Lü[1,3], A. Roy Barman[1,2], Z. Q. Liu[1,2], A. Srivastava[1,2], S. Saha[1,3], Y.L. Zhao[1,2], S.W. Zeng[1,2], S. Dhar[1,3], N. Tuzla[4], E. Olsson[4], Q. Zhang[5,6,7], B. Gu[7,9], S. Yunoki[6,7,8], S. Maekawa[7,9], H. Hilgenkamp[10,11], T. Venkatesan[1,2,3], Ariando[1,2*]

[1]*NUSNNI-Nanocore, National University of Singapore, Singapore 117411, Singapore*

[2]*Department of Physics, National University of Singapore, Singapore 117542, Singapore*

[3]*Department of Electrical and Computer Engineering, National University of Singapore, Singapore 117576, Singapore*

[4]*Department of Applied Physics, Chalmers University of Technology, Göteborg 41296, Sweden*

[5]*Key Laboratory for Advanced Technology in Environmental Protection of Jiangsu Province, Yancheng Institute of Technology, Yancheng, 224051, China*

[6]*Computational Condensed Matter Physics Laboratory, RIKEN ASI, Wako, Saitama 351-0198, Japan*

[7]*CREST, Japan Science and Technology Agency, Kawaguchi, Saitama 332-0012, Japan*

[8]*Computational Materials Science Research Team, RIKEN AICS, Kobe, Hyogo 650-0047, Japan*

[9]*Advanced Science Research Center, Japan Atomic Energy Agency, Tokai 319-1195, Japan*

[10]*MESA+ Institute for Nanotechnology, University of Twente, Enschede 7500 AE, The Netherlands*

[11]*Leiden Institute of Physics, Leiden University, Leiden 2333 CA, The Netherlands*

*Email: ariando@nus.edu.sg

**The observation of a two dimensional electron gas (2DEG)** (*1, 2*)**, superconductivity** (*3, 4*)**, magnetic effects** (*5*) **and electronic phase separation** (*6-8*) **at the interfaces of insulating oxides, especially LaAlO$_3$/SrTiO$_3$, has further enhanced the potential of complex oxides for novel electronics. The occurrence of the 2DEG is strongly believed to be driven by the polarization discontinuity** (*9*) **at the interface between the two oxides. In this scenario, the crystal orientation plays an important role and no conductivity would be expected for** *e.g.*,**  the interface between LaAlO$_3$ and (110)-oriented SrTiO$_3$, which should not have a polarization discontinuity** (*10, 11*)**. Here, we report the observation of unexpected conductivity at the LaAlO$_3$/SrTiO$_3$ interface prepared on (110)-oriented SrTiO$_3$. The conductivity was further found to be strongly anisotropic, with the ratio of the conductance along the different directions parallel to the substrate surface showing a remarkable dependence on the oxygen pressure during deposition. The conductance and its anisotropy are discussed based on the atomic structure at the interface, as revealed by Scanning Transmission Electron Microscopy (STEM) and further supported by density functional theory (DFT) calculations.**

The conductivity in crystalline LaAlO$_3$/SrTiO$_3$ (100) interface is mostly interpreted to originate from the polarization discontinuity (*9*) at the interface between the two oxides. In this case, the ABO$_3$ perovskite structure of the SrTiO$_3$ (100) substrate can be considered as stacks of (SrO)$^0$ and (TiO$_2$)$^0$ (Fig. 1A and 1B), whereas for the LaAlO$_3$ it consists of charged sheets of (LaO)$^{+1}$ and (AlO$_2$)$^{-1}$, which leads to a polarization discontinuity at the interface between the LaAlO$_3$ and SrTiO$_3$ (*9*). On the other hand, for the case of LaAlO$_3$/SrTiO$_3$ interfaces prepared on SrTiO$_3$ (110) substrate (Fig. 1C and 1D), both the SrTiO$_3$ and LaAlO$_3$ can be represented by planar stacks of (ABO)$^{+4}$ and (O$_2$)$^{-4}$ layers as proposed in (*10, 11*), which leads to no polarization discontinuity at the interface. Consequently, no conductivity would be expected for such LaAlO$_3$/SrTiO$_3$ (110) interfaces. Nevertheless, as we report here, a 2DEG also arises at these LaAlO$_3$/SrTiO$_3$ (110) interfaces, with strong anisotropic characteristics.

We prepared LaAlO$_3$/SrTiO$_3$ (110) samples with various LaAlO$_3$ thicknesses ($N$) from 1 to 14 unit cells (uc) under different oxygen partial pressures ($P_{O2}$) ranging from $5\times10^{-5}$ to $5\times10^{-3}$ Torr on atomically flat SrTiO$_3$ (110) substrates. The LaAlO$_3$ films were deposited from a single-crystal LaAlO$_3$ target at 720 °C by pulsed laser deposition. The laser (248 nm) energy density was 1.4 J/cm$^2$ and repetition rate was 1 Hz. During deposition, the film growth was monitored using *in-situ* reflection high energy electron diffraction (RHEED). After deposition, all samples were cooled to room temperature at 10 °C/min in oxygen at the deposition pressure. For comparison, LaAlO$_3$/SrTiO$_3$ (100) samples were grown with 12 uc LaAlO$_3$ layers under the same preparation conditions. Thicker LaAlO$_3$ films of about 15 nm were grown to enable confirmation of the growth orientation of the LaAlO$_3$ on SrTiO$_3$ (110) and (100) substrates by X-ray diffraction (XRD) (supporting online material). The obtained XRD spectra, reflection RHEED-patterns and STEM confirmed the epitaxial growth of LaAlO$_3$ on SrTiO$_3$ (110) (Fig. 1F). For the (100) case, substrates were treated with well-established conditions; buffered HF for 30 sec followed by thermal annealing at 950 °C for 2 hours in air (*12, 13*). For the (110)-oriented SrTiO$_3$, the substrates were annealed for 3 hours at 1000 °C in air (*14*). A detailed Atomic Force Microscopy (AFM) analysis is described in the supporting online material.

The sheet resistance, $R_s$, charge density, $n_s$, and Hall mobility, $\mu$, were measured using a Van der Pauw geometry. Figure 2A shows $R_s$ versus temperature for the LaAlO$_3$/SrTiO$_3$ (110) samples grown at different oxygen partial pressures. Strikingly, these interfaces show conductivity reminiscent to the (100) case (Fig. 2B), albeit with a larger dispersion with respect to the oxygen pressure during growth. The $R_s$ follows a $T^2$-like dependence over a temperature range of 300-100 K and tends to saturate at low temperatures. The samples grown at higher oxygen pressures show a resistance upturn at lower temperatures, a typical localization/Kondo-behavior indicating the presence of magnetic scattering centers at the interface. While Kondo scattering in the (100) case is reported only for LaAlO$_3$ thickness larger than 15 uc (*5, 15, 16*), in the case of (110) it is seen even for 10 uc LaAlO$_3$ thickness indicating a stronger localization. We note here that for the (100) case the dispersion in the resistivity with oxygen deposition pressures was also reported but only for samples with "*thicker*" LaAlO$_3$ layers of 20 uc (*6*), implying different sensitivity to the LaAlO$_3$ thickness between these interfaces. Interestingly, there is a temperature shift in the resistance-minimum with the oxygen growth pressure, indicating that the strength of the localization at the interface is sensitive to the deposition conditions. We note here that resistivity measurements have been carried out down to 2 K, and it is feasible that some of these samples may become superconducting at lower temperatures.

Figure 2C and 2D show the corresponding variation in $n_s$ and $\mu$ with temperature for the LaAlO$_3$/SrTiO$_3$ (110) and (100) samples grown at different oxygen partial pressures. For the (110) samples, $n_s$ decreases with decreasing temperature and for the samples grown at higher partial pressures and $n_s$ at low temperature (2 K) tends towards a constant value of about 1.5×10$^{13}$ cm$^{-2}$, which is somewhat lower than the residual carrier density of about 2.5×10$^{13}$ cm$^{-2}$ for the LaAlO$_3$/SrTiO$_3$ (100) samples which could be due to the preferred occupancy of the Ti d$_{xy}$ orbitals at the interface which localizes electrons (*17, 18*). For the thermally activated charge carriers in the (110)-samples we find an activation energy of about 9 meV, whereas in LaAlO$_3$/SrTiO$_3$ (100) interfaces an activation energy of 6.0 meV has been reported (*19, 20*). For both orientations, the measured $\mu$ follows a power law temperature dependence for $T>100$ K, as $\mu \sim 5\times 10^6 T^{-2.2}$. All the samples grown at higher pressures show $\mu$ values of the order of 200–600 and 2–7 cm$^2$V$^{-1}$s$^{-1}$ at 2 and 300 K, respectively. These values correspond well with those reported in literature for the LaAlO$_3$/SrTiO$_3$ (100) interfaces (*2, 15, 21*). We notice a decrease in the $\mu$ in the (110)-case, simultaneous with Kondo features becoming visible in the $R_s(T)$ curves. For the (110) sample grown at $5\times10^{-5}$ Torr, which has the highest $\mu$ of 12000 cm$^2$V$^{-1}$s$^{-1}$, the $n_s$ at 300 K is $4.9\times10^{14}$ cm$^{-2}$, suggesting contribution from other sources such as oxygen vacancies and defect centers present near the interface.

One of the characteristic features of the LaAlO$_3$/SrTiO$_3$ (100) interfaces, and considered as one of the most important arguments for a polarization discontinuity driven mechanism, is the dependence of the sheet conductivity on the number of unit cells of LaAlO$_3$ with the observation of an insulator to metal transition at 3-4 uc of LaAlO$_3$ layers (*2, 4*). In Fig. 3, the conductivity versus the number of LaAlO$_3$ unit cells $N$ grown on the SrTiO$_3$ (110) substrates is plotted. Here also we clearly see two regimes, an insulating one at low $N$ ($N<3$) and a metallic state at higher $N$ ($N \geq 4$) values.

Considering all the data presented above, it is remarkable that, the (110)-samples are not only conducting, but also bear remarkable similarities to the conductance properties of the (100) case. However, there are also strong differences, the most noticeable of which is the anisotropy of the conductance.

Figure 4A and 4B show the electrical transport characteristics of the LaAlO$_3$/SrTiO$_3$ (110) samples along the [1$\bar{1}$0] and [001] directions, respectively, demonstrating the anisotropic transport behavior. Strikingly, along [1$\bar{1}$0] direction strong localization is seen while along [001] no localization is seen. A basic difference between these two directions is the presence of linear Ti-O-Ti chains along [001] and buckled Ti-O-Ti chains along [1$\bar{1}$0] as depicted schematically in Fig. 4C, and confirmed by STEM study (Fig.1F and supporting online material). The deposition pressure dependence of $R_s$ at 2 K is plotted in Fig. 4D for both directions. Clearly, the conductivity along the [1$\bar{1}$0] direction has a much stronger dependence on deposition pressures compared to the [001] direction indicating the strong role of the differences in the Ti-O-Ti chain along the two orthogonal directions. The anisotropy arising from other effects such as step edges (*22*) has been measured for (100) case and this is substantially smaller than what we see in the (110) case and so we can rule out step edge scattering as the dominant mechanism.

The fact that there are many similar properties to the (100) case, even including the insulator to metal transition at around 4 uc of LaAlO$_3$, indicates that a very similar mechanism is at work for the formation of the 2DEGs for both (100) and (110) cases. The origin for the unexpected conductivity in the (110) case can still also be explained by a modified polarization discontinuity effect, with the interface viewed as depicted in Fig. 1E (buckled interface) rather than the one in Fig. 1C or 1D (planar interface). The electronic configuration of interfaces with such a buckled interface structure as in Fig. 1E was modeled by DFT (supporting online materials). Indeed as shown in Fig. 5, an insulator to metal transition with increasing number of LaAlO$_3$ mono layers is obtained. The critical thickness for insulator to metal transition is about 4 uc which is in agreement with our experimental observation. This result shows that by viewing the LaO/TiO$_2$ as buckled layers, the thickness dependence of the insulator to metal transition arising from polarization catastrophe can be reproduced.

To investigate the atomic structure at the (110) interfaces, we performed cross sectional STEM on the LaAlO$_3$/SrTiO$_3$ (110) interfaces shown in Fig. 1F (supporting online materials). The interface shows perfect epitaxy and the position of the La and Sr atoms indicates that Ti-O-Ti chains in this interface should be buckled as in Fig. 1E. However some of these Ti-O-Ti segments (interpreted by looking at the La and Sr positions) extend up to 2-4 uc rather than 2 uc as anticipated. Both the sparseness of such extended facets and the absence of connectivity among them rule out other mechanisms for conduction (e.g., an effective (100) interface at 45 degrees).

The importance of the (110) interface is better brought out in context with what has been learnt so far with the (100) interfaces. The 2DEG at the LaAlO$_3$/SrTiO$_3$ (100) interfaces has been studied under a variety of deposition conditions, layer thicknesses and externally applied fields (*23-24*). In most of these experiments the 2DEG was presumed to be the result of a direct manipulation of electrostatic potential at the interface in order to promote electrons to the conduction band of SrTiO$_3$, which requires energy of the order of 3.2 eV. However, recently the 2DEG was also demonstrated on the surface of SrTiO$_3$ by shining UV light (*25*), vacuum cleaving (*26*) and chemical reduction of the surface during growth (*27*).

It is clear that the 2DEG can be produced by different mechanisms, which can elevate the electron to the conduction band of SrTiO$_3$. Recent experimental work involving amorphous over layers on SrTiO$_3$ (*27*) showing a conductive interface has brought in the importance of interface chemistry. Under this scenario oxygen diffusion from SrTiO$_3$ surface during film growth into the over layers was deemed responsible for the interface conductivity. What was striking was the similarity of the results between the amorphous and crystalline over layers in terms of the saturation in charge density and conductivity with the layer thickness. Unlike their crystalline counterparts, however, one of the short comings of the amorphous over layers seems to be the thermal stability of the interfaces which lose their conductivity subsequent to a 150 °C anneal in air which will make them unsuitable for device applications. The real attraction of the crystalline interfaces is that they offer a variety of controllable parameters such as strain (*28*), correlation effects (29) and large anisotropy (as shown here) which can be used to manipulate the interface properties.

Our experiment using (110) substrates show that the polarization catastrophe model may still be valid for crystalline $LaAlO_3$/$SrTiO_3$ interfaces provided the interface is viewed appropriately (buckled interface). Because of the different atomic structure in the (110)-samples, the presence of buckled Ti-O-Ti chains and the resultant anisotropy and unusual transport behavior, these (110) interfaces may lead to novel physics and devices.

30. We thank H. Boschker and J. Mannhart for the discussion and the National Research Foundation (NRF) Singapore under the Competitive Research Program (CRP) 'Tailoring Oxide Electronics by Atomic Control' NRF2008NRF-CRP002-024, National University of Singapore (NUS) cross-faculty grant and FRC for the financial support. NT and EO thank the Swedish Research Council and the Area of Advance Nanoscience and Nanotechnology at Chalmers for the financial support.

**Figure Legends**

**Fig. 1.** Representation of LaAlO$_3$/SrTiO$_3$ interface constructed along the (100) and (110) orientations. Layout of the polar catastrophe model for LaAlO$_3$/SrTiO$_3$ interface, on (**A**) (100) and (**C**) (110)-oriented SrTiO$_3$ substrates, where planes are segmented as planar charge sheets. In the case of (100), charge transfer is expected while in the case of (110) there is no polarization discontinuity and hence no charge transfer (*9-11*). (**B**) and (**D**) Atomic picture of the interfaces for representations (**A**) and (**C**), respectively. (**E**) Atomic picture of LaAlO$_3$/SrTiO$_3$ interface on (110)-oriented SrTiO$_3$, considering the (110) planes of SrTiO$_3$ and LaAlO$_3$ as buckled sheets. (**F**) Experimental evidence for such buckling, as revealed by this STEM cross section image (La and Sr atoms marked with colors accordingly).

**Fig. 2.** Transport properties of the LaAlO$_3$/SrTiO$_3$ interfaces grown on (100) and (110)-oriented SrTiO$_3$ substrates. Temperature dependence of the sheet resistance $R_s$ (*T*) of the LaAlO$_3$/SrTiO$_3$ interfaces, for different oxygen partial pressures ($P_{O_2}$) during growth on (**A**) (110) and (**B**) (100)-oriented SrTiO$_3$ substrates. Temperature dependence of carrier density $n_s$ (*T*) and Hall mobility $\mu$ (*T*) of the LaAlO$_3$/SrTiO$_3$ interfaces, for different $P_{O_2}$ during growth on (**C**) (110) and (**D**) (100)-oriented SrTiO$_3$ substrates.

**Fig. 3.** Number of unit cells of LaAlO$_3$ dependence of sheet conductivity for LaAlO$_3$/SrTiO$_3$ (110) interface. The room temperature sheet conductivity as a function of number of unit cells of LaAlO$_3$ for the LaAlO$_3$/SrTiO$_3$ (110) samples, clearly showing the insulator to metal transition (data points marked with open red circle are for a sample initially having 3 uc of LaAlO$_3$, followed by the growth of 2 more uc making it 5 uc in total).

**Fig. 4.** Electrical transport anisotropy of the LaAlO$_3$/SrTiO$_3$ (110) interfaces. $R_s(T)$ measured along (**A**) $[1\bar{1}0]$ and (**B**) [001] directions for the LaAlO$_3$/SrTiO$_3$ (110) samples grown at different oxygen partial pressures. (**C**) Schematic view of The Ti chain arrangement along the $[1\bar{1}0]$ and [001] directions. (**D**) Deposition oxygen pressure dependence of $R_s$ at 2 K measured along the $[1\bar{1}0]$ and [001] directions.

**Fig. 5.** DFT calculations performed on LaAlO$_3$/SrTiO$_3$ (110) structure. (**A**) schematic cell structure of LaAlO$_3$/SrTiO$_3$ (110) interface with TiO terminated SrTiO$_3$ (110), (**B**) The total density of states for different numbers $N$ of LaAlO$_3$ monolayers deposited on SrTiO$_3$ (110) substrate, clearly shows the band gap decrease with increasing $N$ and an insulator to metal transition occurring at 4 uc.

**Annadi *et al.*, Figure 1**

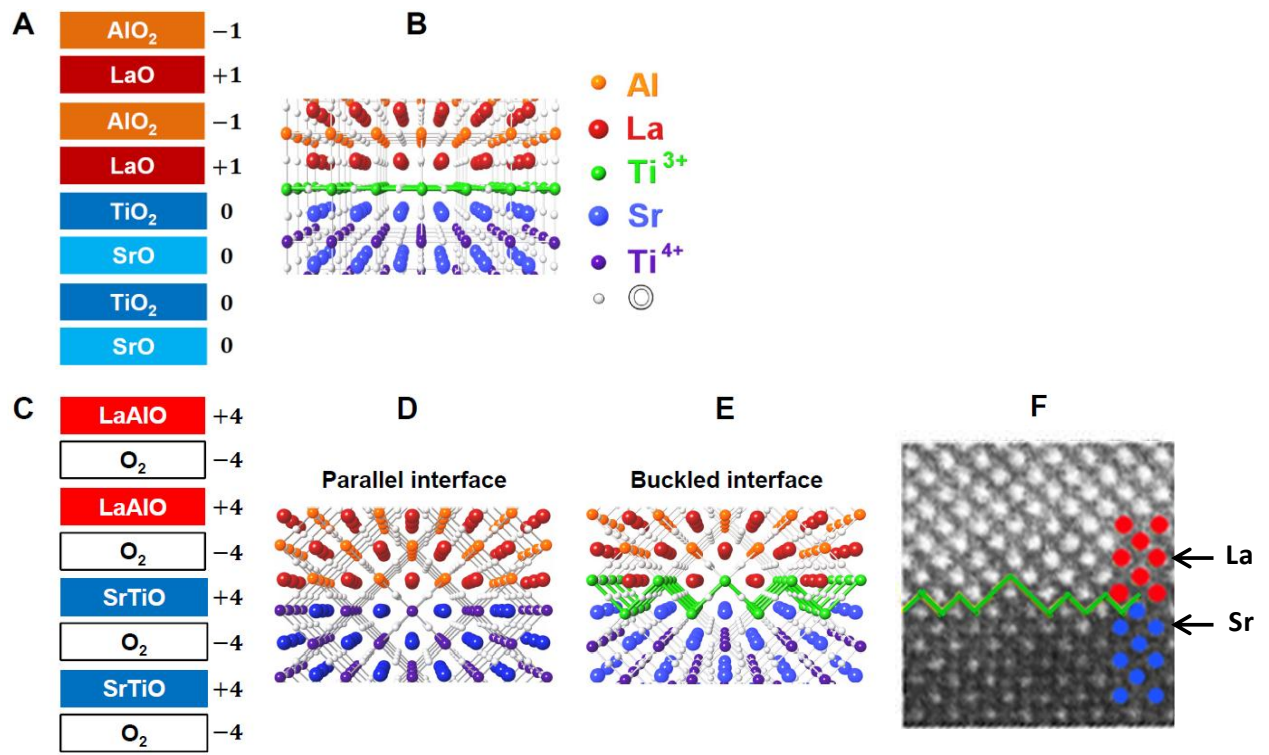

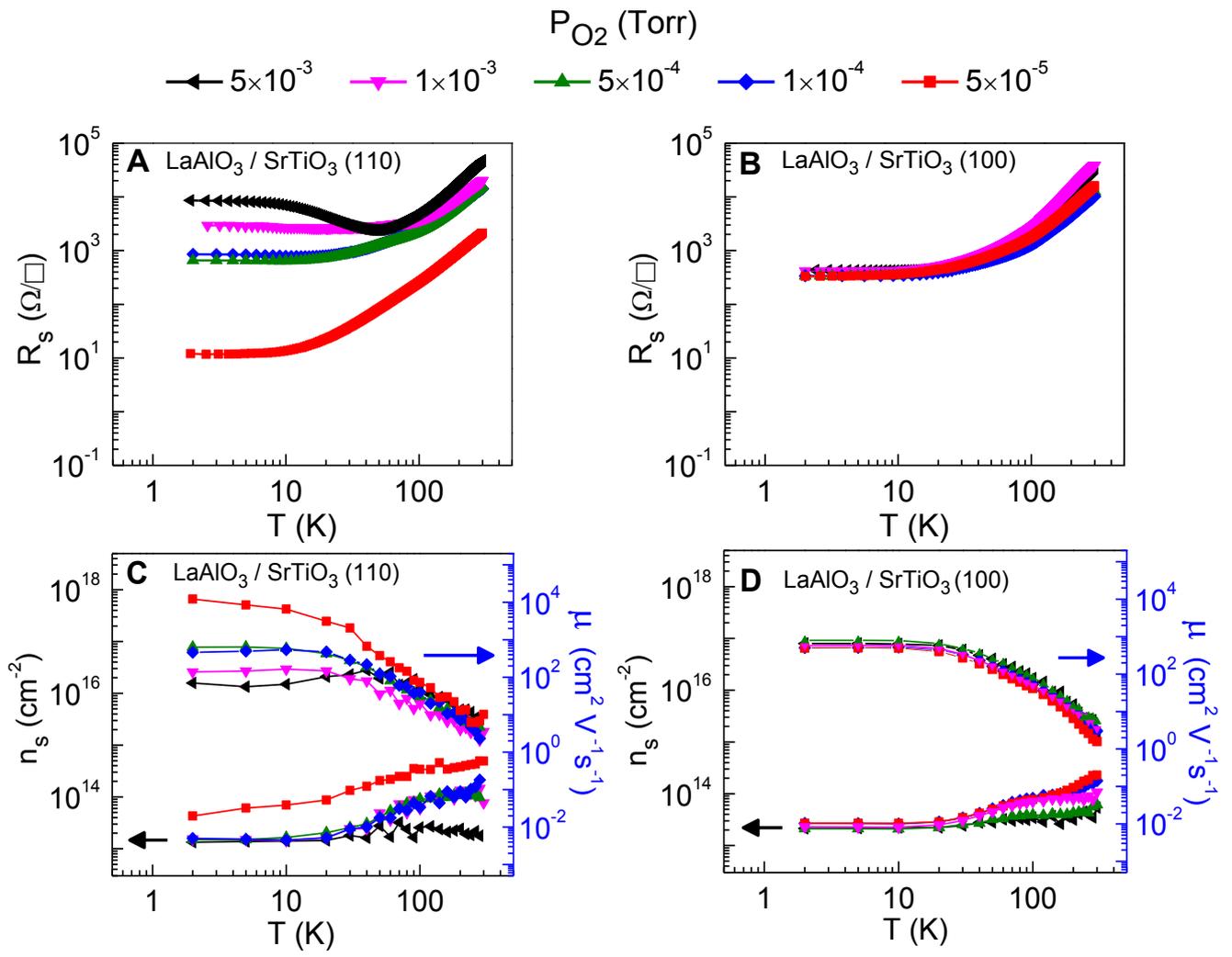

**Annadi *et al.*, Figure 3**

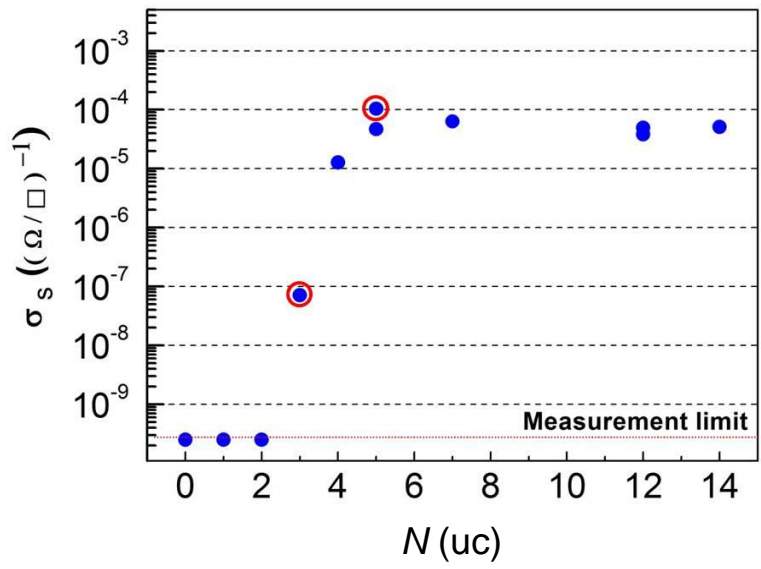

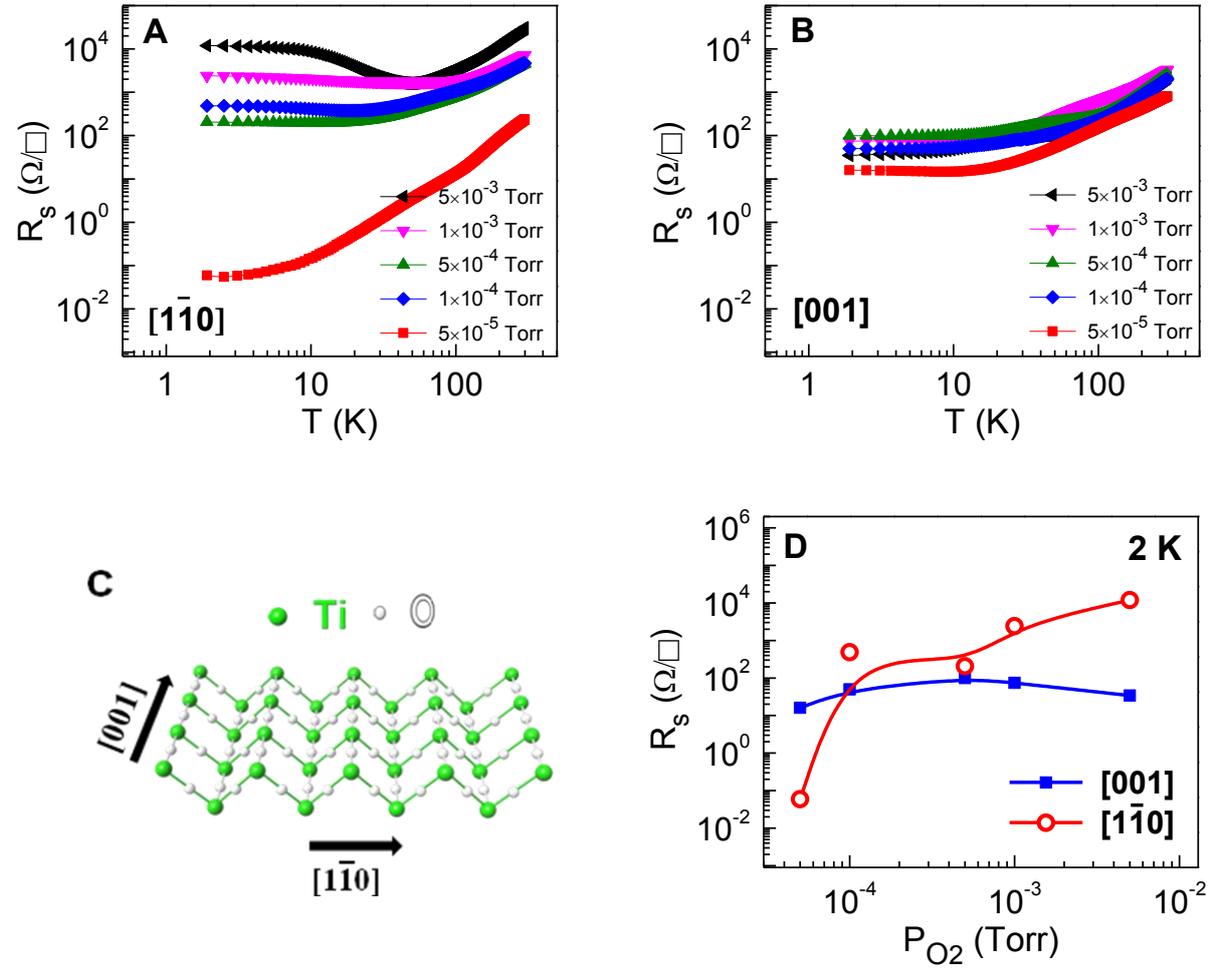

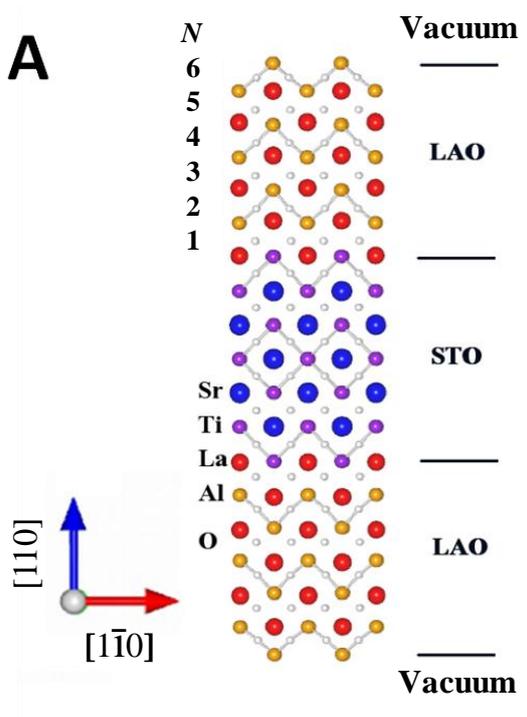
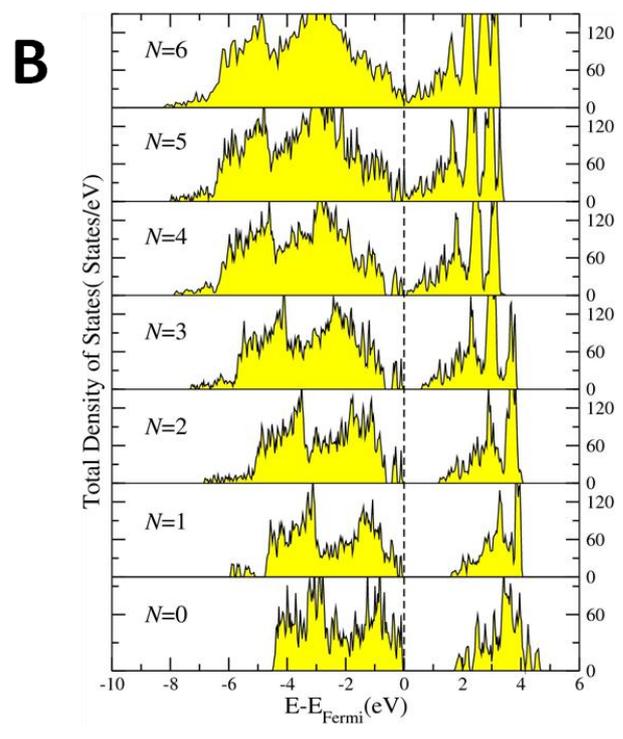